\documentclass{emulateapj}
\usepackage{amsmath}
\usepackage{amssymb}
\usepackage{natbib}
\usepackage{xspace}
\usepackage{graphicx}
\usepackage{rotating} 
\usepackage{float}

\newcommand{\nh}{$N_{\text{H}}$\xspace}

\newcommand{\ka}{K$\alpha$\xspace}

\newcommand{\integral}{\textsl{INTEGRAL}\xspace}

\newcommand{\suzaku}{\textsl{Suzaku}\xspace}
\newcommand{\nustar}{\textsl{NuSTAR}\xspace}

\shorttitle{NGC~4388}
\shortauthors{Kamraj et al.}

\begin{document}
		
\title{The \textsl{NuSTAR} View of the Seyfert 2 Galaxy NGC 4388}

\author{N. Kamraj\altaffilmark{1}, E.~Rivers\altaffilmark{1}, F. A.~Harrison\altaffilmark{1}, M. Brightman\altaffilmark{1}, M.~Balokovi\'c\altaffilmark{1}}

\altaffiltext{1}{Cahill Center for Astronomy and Astrophysics, California Institute of Technology, Pasadena, CA 91125, USA} 
	
\email{Contact: nkamraj@caltech.edu}

\begin{abstract}
	
\noindent We present analysis of \textsl{NuSTAR} X-ray observations in the 3--79 keV energy band of the Seyfert 2 galaxy NGC 4388, taken in 2013. The broadband sensitivity of \nustar, covering the Fe K$\alpha$ line and Compton reflection hump, enables tight constraints to be placed on reflection features in AGN X-ray spectra, thereby providing insight into the geometry of the circumnuclear material. In this observation, we found the X-ray spectrum of NGC 4388 to be well described by a moderately absorbed power law with non-relativistic reflection. We fit the spectrum with phenomenological reflection models and a physical torus model, and find the source to be absorbed by Compton-thin material (N$_{H} = (6.5\pm0.8)\times10^{23}$ cm$^{-2}$) with a very weak Compton reflection hump (R $<$ 0.09) and an exceptionally large Fe K$\alpha$ line (EW $= 368^{+56}_{-53}$ eV) for a source with weak or no reflection. Calculations using a thin-shell approximation for the expected Fe K$\alpha$ EW indicate that an Fe K$\alpha$ line originating from Compton-thin material presents a possible explanation.   

\end{abstract}

	
\section{Introduction}

It is well established that Active Galactic Nuclei (AGN) are powered by accreting supermassive black holes (SMBH), with a significant fraction of AGN prone to moderate to Compton-thick obscuration due to circumnuclear gas and dust. Under the unified model of AGN, much of the obscuration is produced by a uniform, dusty torus, with the observed differences between Seyfert 1s and Seyfert 2s simply arising from different viewing angles of the torus \citep{antonucci,urry-1995}. Several torus spectral models have been developed which self-consistently model toroidal reprocessing of the hard X-ray spectrum, such as the XSPEC models \texttt{MYTorus} \citep{mytorus} and \texttt{BNTorus} \citep{murray}. The picture of a uniform dusty torus is likely an oversimplification of AGN geometry, with a more realistic description being a clumpy torus composed of many optically thick clouds \citep{clumpy-torus-krolik}. Cloud obscuration along the line of sight of the observer smears the rigid border between type 1 and type 2 and thus can lead to misclassification of some objects \citep{unification}. Evidence for a clumpy or filamentary dust structure has been found in mid-infrared observations of the Circinus AGN using the Very Large Telescope Interferometer (VLTI) \citep{tristram-2007}. Such clumpy torus models can also better explain transitions between Sy2 and Sy1 type spectra observed in a number of sources \citep{clumpy-torus-1995}. 

Analysis of the X-ray spectra of Seyfert 2 AGN can provide detailed insight into the geometry of the circumnuclear material. The absorption of soft X-ray photons allows us to measure the column density of the material in the line of sight (\nh) while features such as the Compton reflection hump (CRH) near 20-30 keV and Fe K$\alpha$ line at 6.4 keV produced from reprocessing of the continuum X-ray emission, indicate the global amount of Compton-thick and Compton-thin material, respectively \citep{awaki-1991}. 

NGC 4388 is one of the brightest Seyfert 2s at hard X-ray energies \citep{caballero-2012, baumgartner} with an intrinsic 2--10 keV luminosity L$_{2-10}$ of $\sim4.5\times10^{42}$ erg s$^{-1}$, measured by \citet{masini-2016}. The galaxy is viewed at an inclination angle $i \simeq $ {72\degr} and redshift z $=0.0084$ \citep{philips}, with a SMBH mass of $M_{BH} = (8.5\pm0.2)\times10^{6}$ M$_{\odot}$ determined from water maser measurements \citep{bh-mass-kuo}. 

Several broad-band observations of this source have been performed with X-ray missions such as \textsl{BeppoSAX} \citep{risaliti-2002}, \textsl{INTEGRAL} \citep{fedo} and \suzaku \citep{shirai}. \textsl{INTEGRAL} observations conducted from 2003 to 2009 revealed strong variations in hard X-ray emission in the 20--60 keV band on 3--6 month timescales, while \textsl{RXTE} observations have shown rapid (hour-timescale) variability in the column density of the absorbing medium \citep{risaliti}. Past observations have shown the source to be moderately absorbed, with column densities in the range $10^{23} <$ \nh $<10^{24}$ cm$^{-2}$, thus leading to its classification as a Compton-thin Seyfert 2 galaxy \citep{dewangan-2001,fedo}. Broadband X-ray studies with \textsl{BeppoSAX} and \textsl{INTEGRAL} have been unable to constrain the reflection component in NGC 4388 due to the low sensitivity at energies above 10 keV \citep{shirai}. The \nustar observatory, with its hard X-ray focusing optics, has enabled sensitive broadband observations to be performed and detailed modelling of the CRH and Fe K bandpass, thereby placing tight constraints on reflection features \citep{nustar-harrison}. \citet{masini-2016} studied X-ray absorption in NGC 4388 and other water maser AGN by performing torus model fits to the \nustar data, however these authors did not perform detailed modeling of the reflection and Fe line features.    

In this paper we present an analysis of the hard X-ray spectrum of NGC 4388 from \nustar observations made in 2013 in the 3--79 keV energy range. We investigate both physically motivated torus models and phenomenological ones. We compare our results with previous \textsl{INTEGRAL}, \textsl{Swift}, \suzaku and \textsl{RXTE} measurements. In this work, all uncertainties were calculated at the 90\% confidence level and standard values of the cosmological parameters ($h_{0} = 0.7$, $\Omega_{\Lambda} = 0.7$,  $\Omega_{m} = 0.3$) were used to calculate distances.

\section{Observation and Data Reduction}

The NuSTAR satellite observed NGC 4388 twice on 2013 December 27 with a total exposure time of 22.8 ks. Reduction of raw event data from both modules, FPMA and FPMB \citep{nustar-harrison} was performed using the NuSTAR Data Analysis Software (NuSTARDAS, version 1.2.1), distributed by the NASA High Energy Astrophysics Archive Research Center (HEASARC) within the HEASOFT package, version 6.16. We took instrumental responses from the NuSTAR calibration database (CALDB), version 20150316. Raw event data were cleaned and filtered for South Atlantic Anomaly (SAA) passages using the \texttt{nupipeline} module. We then extracted source energy spectra from the calibrated and cleaned event files using the \texttt{nuproducts} module. Detailed information on these data reduction procedures can be found in the NuSTAR Data Analysis Software Guide \citep{nustardas}. An extraction radius of 60\arcsec\ was used for both the source and background regions. We extracted the background spectrum from source-free regions of the image, and away from the outer edges of the field of view, which have systematically higher background. The spectral files were rebinned using the HEASOFT task \texttt{grppha} to give a minimum of 20 photon counts per bin. We did not include NuSTAR data below 3 keV or above 79 keV. In all our modeling we include a cross-correlation constant of $\sim 1$ between FPMA and FPMB to account for slight differences in calibration \citep{Madsen:2015aa}.

\section{Spectral Modeling}

We performed spectral modeling of NGC 4388 using XSPEC v12.8.2 \citep{xspec}. We used cross sections from \citet{vern} and solar abundances from \citet{wilm}. Fitting a simple power law model showed evidence of a strong Fe K$\alpha$ line and soft absorption, but relatively little Compton reflection (see figure 1 (b)). Adding both an absorbed component with partial covering and a redshifted iron line (XSPEC model \texttt{pcfabs $\times$ (powerlaw $+$ zgauss)}) significantly improved the fit ($\chi^{2}$/dof $=$ 578/535) and showed no evidence of a reflection feature, as seen in the residuals in figure 1 (c). We then applied both a phenomenological slab reflection model and a physically-motivated torus model to the data, as described below:

\begin{table*}
	\centering	
	\caption{Best-fit parameter values for the \texttt{pexrav} \& \texttt{MYTorus} models}
	\renewcommand{\arraystretch}{1.5}
	
	\begin{tabular*}{\textwidth}{@{\extracolsep{\fill} }l c c c c c c c c}
		
		\noalign{\smallskip} \hline \hline \noalign{\smallskip}
		Model & $\chi^{2}$/dof & $\Gamma^{A}$ & PL & Observed & Unabsorbed & \nh & R & EW$_{\rm Fe}$ \\ 
		& & & Norm$^{B}$ & F$_{2-10}^{C}$ & F$_{2-10}^{C}$ & & & \\
		& & & & & & $(10^{23}$ cm$^{-2}$) & & (eV) \\ \hline 
		\\
		\texttt{pexrav} & 577/534 & 1.60$^{+0.11}_{-0.07}$ & 4.65$^{+0.88}_{-0.79}$ & 7.77$^{+0.30}_{-0.26}$ & 22.9$^{+2.3}_{-1.8}$ & 6.5$\pm0.8$ & $<$ 0.09 & 368$^{+56}_{-53}$ \\
		\\
		\texttt{MYTorus} & 576/535 & 1.62$\pm0.08$ & 6.16$^{+2.05}_{-1.49}$ & 7.58$^{+0.17}_{-0.18}$ & 28.1$\pm9.1$ & 5.3$\pm0.7$ & - & 394$^{+131}_{-95}$ \\ 
		\\ \hline
	\end{tabular*}
	\tablecomments{Model parameters from fits to NuSTAR spectral data with the \texttt{pexrav} and \texttt{MYTorus} models. Inclination angle was frozen to {72\degr} for both models. \\ $^{A}$Continuum photon index. \\ $^{B}$Power law normalisation in units of $10^{-3}$ counts s$^{-1}$ keV$^{-1}$ at 1 keV.  \\$^{C}$2--10 keV flux in units of $10^{-12}$ erg cm$^{-2}$ s$^{-1}$.}
	\vspace{5pt}	
\end{table*}

\begin{enumerate}
	
	\item XSPEC model \texttt{pcfabs $\times$ (powerlaw $+$ zgauss $+$ pexrav)}: Models an absorbed power law with a Gaussian Fe K$\alpha$ line and a cold Compton reflection component. \texttt{pcfabs} models absorption with a variable covering factor. \texttt{pexrav} \citep{pexrav} models reflection off a slab of infinite extent and optical depth covering between 0 and 2$\pi$ Sr of the sky relative to the illuminating source, corresponding to R between 0 and 1. 
	
	\item XSPEC model \texttt{MYTorus} \citep{mytorus}: Obscuring material is arranged uniformly in a toroidal structure around the central AGN, with a fixed opening angle of {60\degr}. Provides self-consistent modeling of the scattered power law, Fe K$\alpha$ line, and Compton reflection features.  
	
\end{enumerate}

\begin{figure}[t]
	%\centering
	\hspace{-8pt}
	\includegraphics[width=0.5\textwidth]{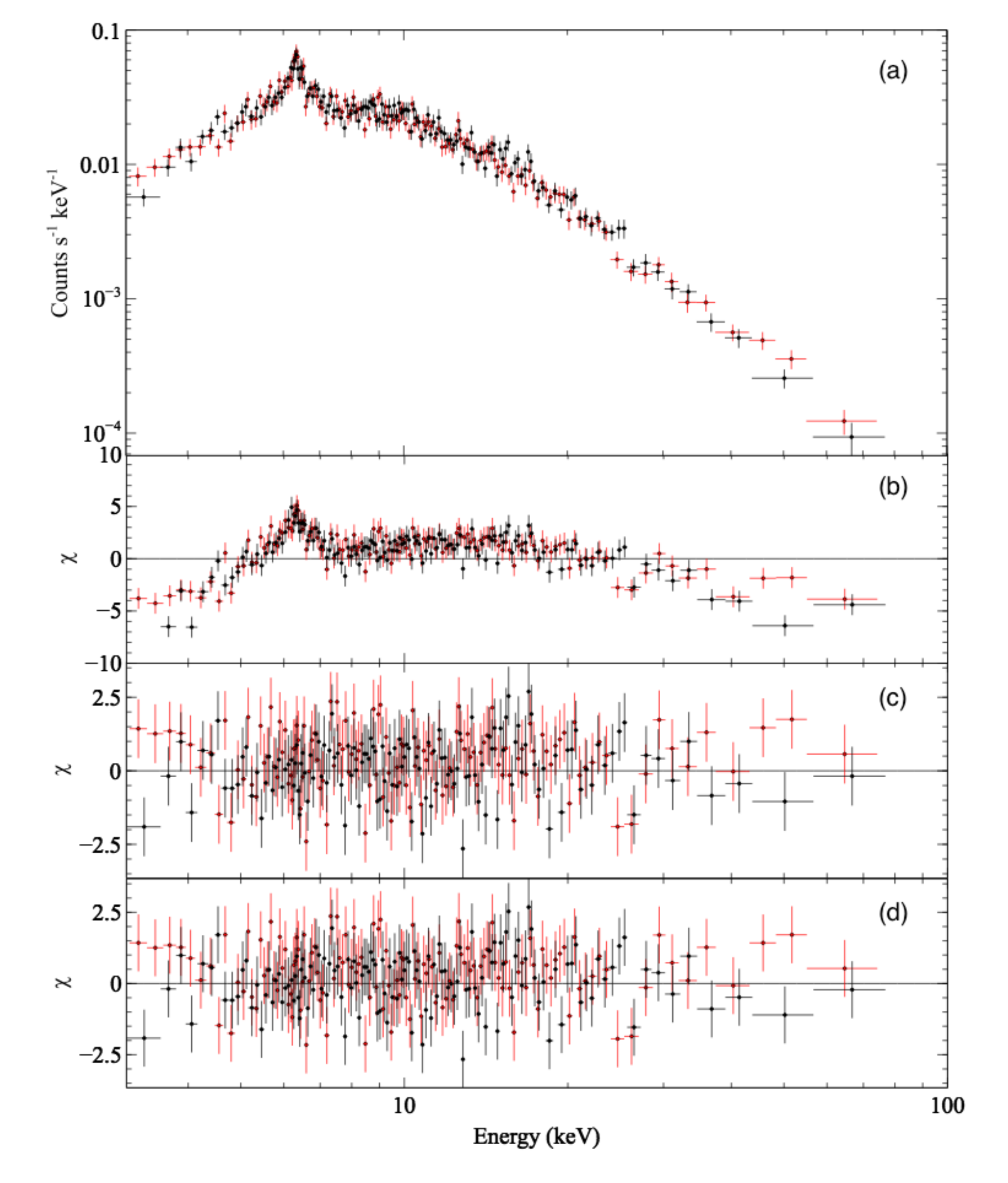}
	\label{fig1}
	\vspace{-8pt}
	\caption{Folded \textsl{NuSTAR} energy spectrum of NGC 4388 (a) alongside fit residuals for: (b) a simple power law model, (c) absorbed power law with an Fe \ka line, (d) absorbed power law with Fe \ka line and a reflection component incorporated using the \texttt{pexrav} model. Black points correspond to FPMA data while points in red correspond to FPMB.}
	\vspace{10pt}
\end{figure}

\begin{figure}[t]
	\centering
	\hspace{-14pt}
	\includegraphics[width=0.5\textwidth]{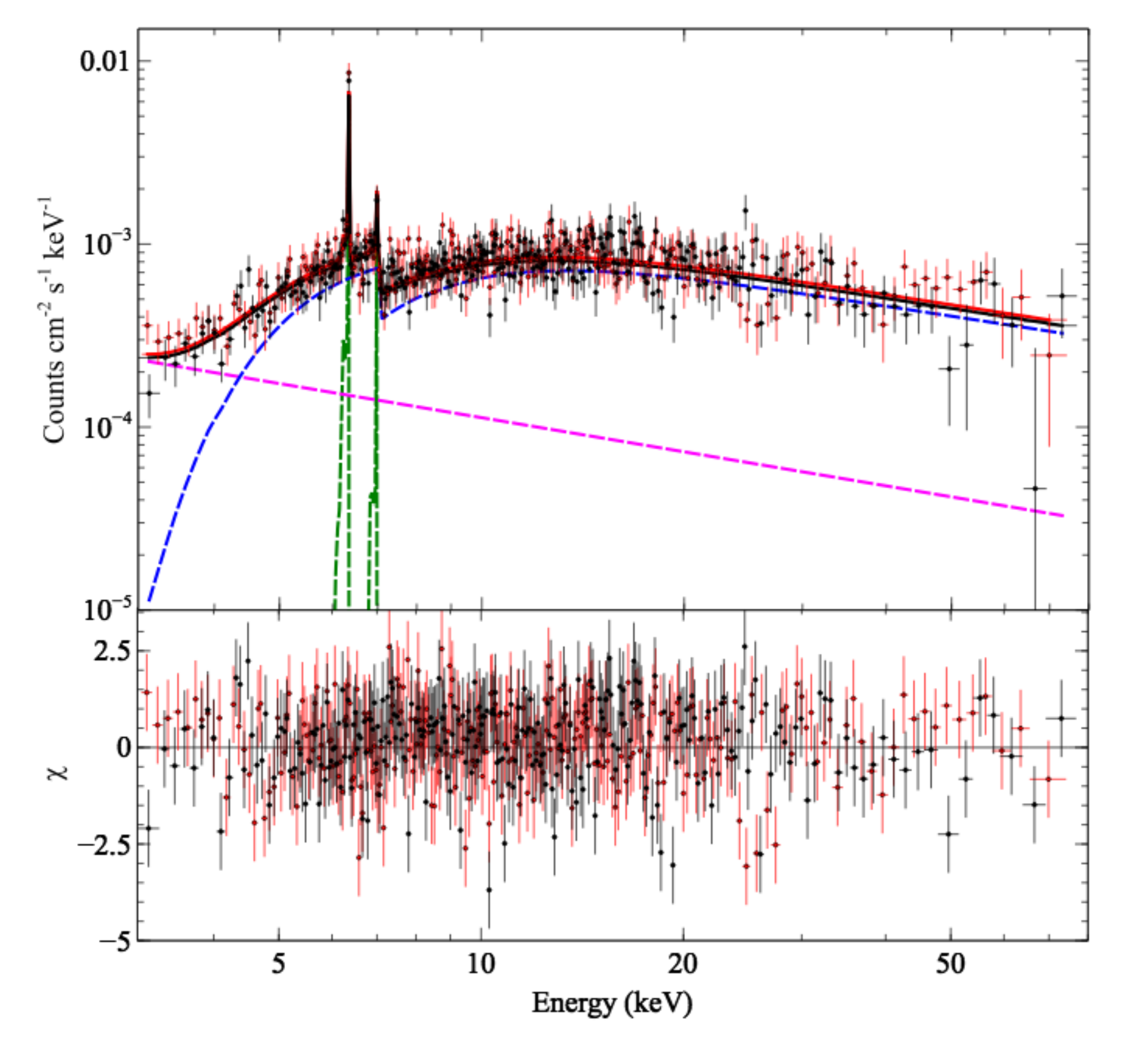} 
	\label{fig2}
	%\vspace{-5pt}
	\caption{\emph{Upper panel}: Unfolded \textsl{NuSTAR} energy spectrum with the \texttt{MYTorus} model fit. Solid lines refer to total model while dashed lines depict individual model components. Model components consist of: power law continuum (magenta), Fe line complex (green) and absorbed power law (blue). \emph{Lower panel}: residuals for the \texttt{MYTorus} best fit model. Data points in black and red correspond to FPMA and FPMB observations respectively.}
	\vspace{8pt}
\end{figure}

Phenomenological models have some limitations in that the reflection component, e.g. modelled via \texttt{pexrav}, is produced from a slab of infinite extent and optical depth. \texttt{MYTorus} provides a more realistic description, by modeling the obscuring material in a toroidal geometry with a finite optical depth. However, \texttt{MYTorus} is limited in that it assumes a uniform density torus, with a sharp change in the line of sight absorption at the edges of the torus. 

For the \texttt{pexrav} model, we set iron and light element abundances to solar, the cutoff energy $E_{c}$ was fixed to 1000 keV and the inclination angle of the plane of reflecting material was fixed to the inclination of the galaxy ({72\degr}). Inclination values were found to be unconstrained when left free, thus justifying using a fixed value for this parameter. The normalisation and photon index of the incident power law were tied to those of the reflected continuum power law. 

For the \texttt{MYTorus} model, we performed fits with the torus inclination fixed at {72\degr}, matching the observed inclination of the galaxy. Column densities, photon indices and inclinations of the scattered continuum and emission line were tied to those of the zeroth-order continuum. The normalisation of the scattered continuum was tied to the zeroth-order continuum whilst the line normalisation was left free. We found that coupling the continuum, reflection and Fe line regions in this manner provided the best fit to the data. 

Table 1 shows the best fit parameter values for both the phenomenological \texttt{pexrav} model and physically motivated \texttt{MYTorus} model.  
Figure 1 shows the corresponding data and residuals for the \texttt{pexrav} model fit; the unfolded energy spectrum is shown with model components and residuals for the \texttt{MYTorus} model in figure 2. 
If we assume reflection off a centrally illuminated, Compton-thick disk with solar abundances, then the expected EW of the Fe line with respect to the flux of the CRH is $<$10 eV, as detailed in \citet{fabian}. 
 
We found NGC 4388 to be Compton-thin with a very weak CRH (R $<$ 0.09), yet it exhibits a large Fe K$\alpha$ line EW. To explain such a high Fe line EW through a large Fe abundance would require unphysically high supersolar iron abundance values, thus we rule out such a scenario. The best fit Fe K$\alpha$ line width is consistent with zero (upper limit of $<$ 70 eV), indicating an absence of line broadening and implying that the fluorescent material is located far from the central source. We found a covering factor of 0.9 for the absorbing material, pointing to a small opening angle of the torus. Values of $\Gamma$, \nh, F$_{2-10}$ and EW$_{Fe}$ are consistent between the two models applied, with both models producing equally good fits to the data, indicating that physically motivated torus models provide a statistically equivalent description of the hard X-ray spectra compared to phenomenological slab models, and are suitable for characterising this Compton-thin source. 

As a check that our spectral modeling succesfully recovers intrinsic AGN parameters, we calculated the Eddington ratio $\lambda_{Edd} = L_{Bol}/L_{Edd}$ using a bolometric correction of 10 to the intrinsic 2--10 keV luminosity. Given a black hole mass of 8.5$\times10^{6}$ M$_{\odot}$, $L_{Edd} \simeq 1.1\times10^{45}$ erg s$^{-1}$. We found the intrinsic 2--10 keV luminosity to be $\sim 8.9\times10^{41}$ erg s$^{-1}$. The expected value of $\Gamma$ can be calculated from the known relationship between $\Gamma$ and $\lambda_{Edd}$, as detailed in \citet{gamma-edd-ratio}, and was found to be $\sim$ 1.60, which is in good agreement with the best fit $\Gamma$ values from our spectral modeling.

\section{Discussion}

Overall, the \textsl{NuSTAR} data shows that the spectrum of NGC 4388 is well characterised by both phenomenological models and physically motivated torus models. We found the source to be moderately obscured by Compton-thin material, with a very weak reflection component and a strong Fe \ka line. The best fit results for $\Gamma$ and \nh were in fairly good agreement with past observations by \textsl{INTEGRAL} \citep{fedo,beckmann} and \textsl{Swift} \citep{fedo}. The constraint on the reflection component found from the \texttt{pexrav} model of R $< 0.09$ is consistent with that obtained from archival \textsl{RXTE} data \citep{agn-rossi-rivers,risaliti} and combined long-term \textsl{Swift} \& \textsl{INTEGRAL} observations \citep{fedo}.  

The lack of Compton reflection in NGC~4388 is not unusual. \citet{agn-rossi-rivers} found that 15\% of Seyfert-like AGN in the \textsl{RXTE} archive show no significant contribution from a CRH ($R < 0.1$), however three of the five sources (Cen~A, Cyg~A, and 3C~111) are actually radio loud AGN and all show weak Fe lines as well.  A handful of other Compton-thin Seyfert 2's in the sample had large Fe lines
with little contribution from Compton reflection, but lacked the statistics to place good constraints on $R$. One of these, NGC~2110, was observed by \textsl{Suzaku} and analyzed by \citet{Rivers:2014aa}, which 
confirmed the lack of reflection, but found much lower values for the Fe line EW as well ($\sim$ 50 eV).  

One possibility that is investigated by \citet{Rivers:2014aa} is a scenario in which there is a large global amount of material that is not Compton-thick. To calculate the expected amount of Fe emission from Compton-thin material, we can use a thin-shell approximation and the following equation based on \citet{EW-NH-calc}:

\begin{center}
\begin{equation}  EW_{\rm K\alpha} = f_{\rm c}\, \omega\, f_{\rm K\alpha}\, A_{\rm abund}\, N_\text{H}\, \frac{\int_{E_{\rm K-edge}}^{\infty} P(E)\, \sigma_{\rm ph}(E)\, {\rm d}E}{P(E_{\rm line})}  \end{equation}
\end{center}

with $f_{\rm c}$ the covering fraction of the absorber, $\omega$ the fluorescent yield, $f_{\rm K\alpha}$ the fraction of photons contributing to the Fe \ka line production, $A_{\rm abund}$ the Fe abundance relative to hydrogen, $P$(E) the continuum power law, and $\sigma_{\rm ph}(E)$ the K-shell absorption cross section as a function of energy.  
Assuming solar abundances for $A_{\rm abund}$ and using values for the fluorescent yield and cross section from \citet{EW-NH-calc},  
we can calculate the contribution to the Fe line EW from a uniform shell of material with the column density given in Table 1 for the line of sight absorption, that is centred on a continuum emission source which is assumed to be an isotropic, point-source emitter.

We find a maximum EW of 548 eV assuming a 100\% sky-covering fraction, which is consistent with our measured EW, indicating that an Fe line originating from Compton-thin material could be a plausible scenario. Previous studies of NGC 4388 \citep{beckmann} have also suggested that the absence of a CRH in the hard X-ray region may point to non-isotropic emission of radiation that fails to illuminate the disk. Another possibility is a poorly illuminated torus with a very large opening angle (that is, a very flat torus), or a more complex geometry of the circumnuclear material \citep{fabian}. It is worth noting that our best-fit \texttt{MYTorus} model physically corresponds to a Compton-thin torus intersecting the line of sight and also providing the necessary Fe line flux.

While our analysis of the \nustar observation of NGC 4388 found the CRH to be absent in this source, the detection of a CRH has been reported in a past observation of NGC 4388 with \suzaku. \citet{shirai} analyzed a 100 ks \suzaku observation of NGC~4388 from 2005 December, and for the first time were able to detect the CRH (R $= 1.40^{+0.29}_{-0.36}$). They postulated that since the source had decreased in luminosity compared to previous \integral observations reported in \citet{beckmann}, they could be seeing a delay in the decrease of the CRH due to a light echo in Compton-thin material that is light years away from the central source. The Fe \ka line was resolved in the XIS data with a width of $45^{+5}_{-6}$ eV from simultaneous broadband fits of XIS and HXD data, corresponding to a radial distance of $\sim 0.01$ pc. However, the paper concluded that the Fe \ka line broadening was attributed to the Compton shoulder rather than intrinsic broadening from material close to the central source. Furthermore, the \suzaku observation revealed short term flux variablility (half-day timescales) wherein the reflection fraction R changes but the reflection flux and Fe \ka line flux did not vary significantly. Thus the fluorescent iron line emission appears to be decoupled from direct emission and likely originates in distant reflecting material located several light years from the continuum source, consistent with past \textsl{INTEGRAL} observations \citep{beckmann}. The 2 - 10 keV flux observed by \suzaku was found to be $2.0\times10^{-11}$ erg cm $^{-2}$ s$^{-1}$, which is the same order of magnitude as that observed with \nustar ($\sim8.0\times10^{-12}$ erg cm$^{-2}$ s$^{-1}$), indicating it is likely that the source remained in a relatively low flux state over long timescales and thus the CRH disappeared from the data in the 7 years between the \suzaku and \nustar observations. 

\section{Summary}

Our spectroscopic analysis of the moderately obscured Seyfert 2 galaxy NGC 4388 from \nustar observations revealed the hard X-ray spectrum to be well represented by both phenomenological reflection models and physically motivated torus models. One possible explanation for the exceptionally large EW of the Fe \ka line and weak CRH in this source is the presence of a large global amount of material which is capable of producing the observed EW but is not sufficiently thick to produce a distinguishable CRH feature. The detection of a CRH in the X-ray spectrum of NGC 4388 from \suzaku observations performed in 2005 can be explained by a light echo in Compton-thin material located light years from the central source, resulting from the source being in a low flux state several years prior to the \nustar observation. However, further multi-epoch hard X-ray monitoring will be needed to conclude whether this is a likely explanation.

\vspace{10pt}

%The authors thank the anonymous referee for helpful and constructive comments contributing to this work. This work was supported under NASA Contract No. -- and sub-contract No. --. 

We have made use of data from the NuSTAR mission, a project led by the California Institute of Technology, managed by the Jet Propulsion Laboratory, and funded by the National Aeronautics and Space Administration. We thank the NuSTAR Operations, Software and Calibration teams for support with the execution and analysis of these observations. This research has made use of the NuSTAR Data Analysis Software (NuSTARDAS) jointly developed by the ASI Science Data Center (ASDC, Italy) and the California Institute of Technology (USA). M. Balokovi\'c acknowledges support from NASA Headquarters under the NASA Earth and Space Science Fellowship Program, grant NNX14AQ07H. 

\facility{\emph{Facility}: \nustar}

%\pagebreak 

\bibliographystyle{apj}
\maketitle
\bibliography{ngc4388}

\begin{thebibliography}{}
\expandafter\ifx\csname natexlab\endcsname\relax\def\natexlab#1{#1}\fi

\bibitem[{Antonucci(1993)}]{antonucci}
Antonucci, R. 1993, \aap, 31, 473

\bibitem[{{Aretxaga} {et~al.}(1999){Aretxaga}, {Joguet}, {Kunth}, {Melnick}, \&
  {Terlevich}}]{clumpy-torus-1995}
{Aretxaga}, I., {Joguet}, B., {Kunth}, D., {Melnick}, J., \& {Terlevich}, R.~J.
  1999, \apjl, 519, L123

\bibitem[{Arnaud(1996)}]{xspec}
Arnaud, K. 1996, in ASP Conf. Series, Vol. 101, Astronomical Data Analysis
  Software and Systems, ed. G.~H. {Jacoby} \& J.~{Barnes}

\bibitem[{{Awaki} {et~al.}(1991){Awaki}, {Kunieda}, {Tawara}, \&
  {Koyama}}]{awaki-1991}
{Awaki}, H., {Kunieda}, H., {Tawara}, Y., \& {Koyama}, K. 1991, \pasj, 43, L37

\bibitem[{{Baumgartner} {et~al.}(2013){Baumgartner}, {Tueller}, {Markwardt},
  {Skinner}, {Barthelmy}, {Mushotzky}, {Evans}, \& {Gehrels}}]{baumgartner}
{Baumgartner}, W.~H., {Tueller}, J., {Markwardt}, C.~B., {et~al.} 2013, \apjs,
  207, 19

\bibitem[{Beckmann {et~al.}(2004)Beckmann, Gehrels, Favre, Walter, Courvoisier,
  Petrucci, \& Malzac}]{beckmann}
Beckmann, V., Gehrels, N., Favre, P., {et~al.} 2004, ApJ, 614, 641

\bibitem[{Brightman \& Nandra(2011)}]{murray}
Brightman, M., \& Nandra, K. 2011, \mnras, 413, 1206

\bibitem[{{Brightman} {et~al.}(2016){Brightman}, {Masini}, {Ballantyne},
  {Balokovi{\'c}}, {Brandt}, {Chen}, {Comastri}, {Farrah}, {Gandhi},
  {Harrison}, {Ricci}, {Stern}, \& {Walton}}]{gamma-edd-ratio}
{Brightman}, M., {Masini}, A., {Ballantyne}, D.~R., {et~al.} 2016, ApJ, 826, 93

\bibitem[{{Caballero-Garcia} {et~al.}(2012){Caballero-Garcia}, {Papadakis},
  {Nicastro}, \& {Ajello}}]{caballero-2012}
{Caballero-Garcia}, M.~D., {Papadakis}, I.~E., {Nicastro}, F., \& {Ajello}, M.
  2012, \aap, 537, A87

\bibitem[{{Dewangan}(2001)}]{dewangan-2001}
{Dewangan}, G.~C. 2001, Bulletin of the Astronomical Society of India, 29, 463

\bibitem[{Elitzur(2012)}]{unification}
Elitzur, M. 2012, \apjl, 747, L33

\bibitem[{Elvis {et~al.}(2004)Elvis, Risaliti, Nicastro, Miller, Fiore, \&
  Puccetti}]{risaliti}
Elvis, M., Risaliti, G., Nicastro, F., {et~al.} 2004, ApJ, 615, L25

\bibitem[{Fedorova {et~al.}(2011)Fedorova, Beckmann, Neronov, \& Soldi}]{fedo}
Fedorova, E.~V., Beckmann, V., Neronov, A., \& Soldi, S. 2011, \mnras, 417,
  1140

\bibitem[{George \& Fabian(1991)}]{fabian}
George, I.~M., \& Fabian, A. 1991, \mnras, 249, 352

\bibitem[{{Harrison} {et~al.}(2013){Harrison}, {Craig}, {Christensen},
  {Hailey}, {Zhang}, {Boggs}, {Stern}, {Cook}, {Forster}, {Giommi},
  {Grefenstette}, {Kim}, {Kitaguchi}, {Koglin}, {Madsen}, {Mao}, {Miyasaka},
  {Mori}, {Perri}, {Pivovaroff}, {Puccetti}, {Rana}, {Westergaard}, {Willis},
  {Zoglauer}, {An}, {Bachetti}, {Barri{\`e}re}, {Bellm}, {Bhalerao},
  {Brejnholt}, {Fuerst}, {Liebe}, {Markwardt}, {Nynka}, {Vogel}, {Walton},
  {Wik}, {Alexander}, {Cominsky}, {Hornschemeier}, {Hornstrup}, {Kaspi},
  {Madejski}, {Matt}, {Molendi}, {Smith}, {Tomsick}, {Ajello}, {Ballantyne},
  {Balokovi{\'c}}, {Barret}, {Bauer}, {Blandford}, {Brandt}, {Brenneman},
  {Chiang}, {Chakrabarty}, {Chenevez}, {Comastri}, {Dufour}, {Elvis}, {Fabian},
  {Farrah}, {Fryer}, {Gotthelf}, {Grindlay}, {Helfand}, {Krivonos}, {Meier},
  {Miller}, {Natalucci}, {Ogle}, {Ofek}, {Ptak}, {Reynolds}, {Rigby},
  {Tagliaferri}, {Thorsett}, {Treister}, \& {Urry}}]{nustar-harrison}
{Harrison}, F.~A., {Craig}, W.~W., {Christensen}, F.~E., {et~al.} 2013, ApJ,
  770, 103

\bibitem[{{Krolik} \& {Begelman}(1988)}]{clumpy-torus-krolik}
{Krolik}, J.~H., \& {Begelman}, M.~C. 1988, \apj, 329, 702

\bibitem[{Kuo {et~al.}(2011)Kuo, Braatz, Condon, Impellizzeri, Lo, Zaw,
  Schenker, Henkel, Reid, \& Greene}]{bh-mass-kuo}
Kuo, C.~Y., Braatz, J.~A., Condon, J.~J., {et~al.} 2011, ApJ, 727, 20

\bibitem[{Madsen {et~al.}(2015)Madsen, Harrison, Markwardt, An, Grefenstette,
  Bachetti, Miyasaki, Kitaguchi, Bhalerao, Boggs, Christensen, Craig, Forster,
  Fuerst, Hailey, Perri, Puccetti, Rana, Stern, Walton, Westergaard, \&
  Zhang}]{Madsen:2015aa}
Madsen, K.~K., Harrison, F.~A., Markwardt, C.~B., {et~al.} 2015, ApJ, 220, 8

\bibitem[{Magdziarz \& Zdziarski(1995)}]{pexrav}
Magdziarz, P., \& Zdziarski, A.~A. 1995, MNRAS, 273, 837

\bibitem[{{Masini} {et~al.}(2016){Masini}, {Comastri}, {Balokovi{\'c}}, {Zaw},
  {Puccetti}, {Ballantyne}, {Bauer}, {Boggs}, {Brandt}, {Brightman},
  {Christensen}, {Craig}, {Gandhi}, {Hailey}, {Harrison}, {Koss}, {Madejski},
  {Ricci}, {Rivers}, {Stern}, \& {Zhang}}]{masini-2016}
{Masini}, A., {Comastri}, A., {Balokovi{\'c}}, M., {et~al.} 2016, \aap, 589,
  A59

\bibitem[{Murphy \& Yaqoob(2009)}]{mytorus}
Murphy, K.~D., \& Yaqoob, T. 2009, MNRAS, 397, 1549

\bibitem[{Perri {et~al.}(2014)Perri, Puccetti, Spagnuolo, Davis, Forster,
  Grefenstette, Harrison, \& Madsen}]{nustardas}
Perri, M., Puccetti, S., Spagnuolo, N., {et~al.} 2014, The NuSTAR Data Analysis
  Software Guide v1.7

\bibitem[{Philips \& Malin(1982)}]{philips}
Philips, M.~M., \& Malin, D.~F. 1982, MNRAS, 199

\bibitem[{Risaliti(2002)}]{risaliti-2002}
Risaliti, G. 2002, A\&A, 386, 379

\bibitem[{Rivers {et~al.}(2013)Rivers, Markowitz, \&
  Rothschild}]{agn-rossi-rivers}
Rivers, E., Markowitz, A., \& Rothschild, R. 2013, ApJ, 772, 114

\bibitem[{Rivers {et~al.}(2014)Rivers, Markowitz, Rothschild, Bamba, Fukazawa,
  Okajima, Reeves, Terashima, \& Ueda}]{Rivers:2014aa}
Rivers, E., Markowitz, A., Rothschild, R., {et~al.} 2014, ApJ, 786, 126

\bibitem[{Shirai {et~al.}(2008)Shirai, Fukazawa, Sasada, Ohno, Yonetoku,
  Yokota, Fujimoto, Murakami, Terashima, Awaki, \& Ikeda}]{shirai}
Shirai, H., Fukazawa, Y., Sasada, M., {et~al.} 2008, \pasj, 60, 263

\bibitem[{{Tristram} {et~al.}(2007){Tristram}, {Meisenheimer}, {Jaffe},
  {Schartmann}, {Rix}, {Leinert}, {Morel}, {Wittkowski}, {R{\"o}ttgering},
  {Perrin}, {Lopez}, {Raban}, {Cotton}, {Graser}, {Paresce}, \&
  {Henning}}]{tristram-2007}
{Tristram}, K.~R.~W., {Meisenheimer}, K., {Jaffe}, W., {et~al.} 2007, \aap,
  474, 837

\bibitem[{{Urry} \& {Padovani}(1995)}]{urry-1995}
{Urry}, C.~M., \& {Padovani}, P. 1995, \pasp, 107, 83

\bibitem[{Verner {et~al.}(1996)Verner, Ferland, Korista, \& Yakovlev}]{vern}
Verner, D.~A., Ferland, G.~J., Korista, K.~T., \& Yakovlev, D.~G. 1996, ApJ,
  465, 487

\bibitem[{Wilms {et~al.}(2000)Wilms, Allen, \& McCray}]{wilm}
Wilms, J., Allen, A., \& McCray, R. 2000, ApJ, 542, 914

\bibitem[{Yaqoob {et~al.}(2001)Yaqoob, George, Nandra, Turner, Serlemitsos, \&
  Mushotzky}]{EW-NH-calc}
Yaqoob, T., George, I.~M., Nandra, K., {et~al.} 2001, ApJ, 546, 2

\end{thebibliography}

\end{document}